\documentclass[twocolumn,prl,aps]{revtex4}
\newcommand{\nd}{\noindent}
\newcommand{\be}{\begin{equation}}
\newcommand{\ee}{\end{equation}}
\newcommand{\ben}{\begin{eqnarray}}
\newcommand{\ba}{\begin{eqnarray}}
\newcommand{\een}{\end{eqnarray}}
\newcommand{\ea}{\end{eqnarray}}

\draft

\begin{document}
\title{Thermodynamics' first law: what information theory tells us}
\author{S. Mart\'{\i}nez${^{1,\,2}}$\thanks{%
E-mail: martinez@venus.fisica.unlp.edu.ar},  A.
Plastino${^{1,\,2}}$%
\thanks{
E-mail: plastino@venus.fisica.unlp.edu.ar}, and B. H. Soffer$^3$}
\address{$1$ Instituto de F\'{\i}sica de La Plata,
National University La Plata, C.C.%
727 , 1900 La Plata, Argentina. \\ $^2$ Argentine National
  Research Council
(CONICET) \\$^3$ University of California at Los Angeles, Department
of Electrical Engineering \\ Postal address: 665 Bienveneda Avenue,
Pacific Palisades, California 90272}


\begin{abstract}

\nd Thermodynamics, and in particular its first law, is of
fundamental importance to Science, and therefore of great general
interest to all physicists.  The first law, although undoubtedly
true, and believed by everyone to be true because of its many
verified consequences, rests on a rather weak experimental {\it
foundation} as its path independent aspect has never been {\it
directly} verified, and rests on a somewhat weak foundation
apropos the need for invoking the so-called adiabatic theorem (AT)
to prove it from first principles. We provide here a more direct
and convincing theoretical demonstration, without the AT and some
other usually employed axioms.


\end{abstract}

\maketitle

\section{Introduction}

\nd Jaynes' pioneering 1957 papers in The Physical Review
\cite{jaynes} constitute a new theoretical foundation for the
development of statistical mechanics providing us with a solid
alternative
 on the basis of information theory (IT)
\cite{jaynes2}. One of the main ingredients in Jaynes's treatment
was an intensive use of the principle of parsimony (PP) \cite{katz}.
Our purpose here is to present an original derivation of
thermodynamics' first law from an IT viewpoint and, at the same
time, provide a pedagogical example illustrating  the principle of
parsimony that explicitly invokes Jaynes' discovery (of more than
forty years ago) \cite{jaynes,jaynes2}, that no reference to
equilibrium needs to be made in order to deal completely and
successfully with all of thermodynamics. To some physicists, even
theoreticians, the idea sounds revolutionary even today!

\nd The PP, or Ockham's razor, is a basic methodological principle
that governs scientific endeavor \cite{oxford}. It dictates
simplicity in theory construction, as for instance, in the number
of axioms, or of parameters, involved in a theoretical construct.

\nd As stated above, we will here apply this principle with
regards to the usual IT-treatment of the first law of the
thermodynamics \cite{katz}. Appeal to Ockham's razor will yield a
simpler derivation than the usual text-book one.

\section{Jaynes' approach and Ockham's razor}

\nd The orthodox formulation of statistical mechanics is due to
Gibbs' \cite{gibbs}, working on a classical mechanics substratum.
It is based upon the following set of axioms \cite{marge}

\begin{enumerate}
 \item Ensemble postulate: the system at equilibrium can be represented
by an
 appropriately designed ensemble. \item Equal a priori
probabilities  for cells in phase space. \item The phase space
probability distribution depends {\it only} on the system's
Hamiltonian.
\item This dependence is of {\it exponential} form.
\end{enumerate}
 Jaynes reformulated statistical mechanics in 1957 \cite{jaynes}
 by recourse to information theory concepts \cite{katz}, with quantum
 mechanics now providing
 the background. Instead of a distribution
 function in phase space we now use  a density operator  $\hat \rho$
 to describe our system.  $\hat \rho$ is obtained
  via the so called MaxEnt principle, namely,
  the constrained (Lagrange) maximization of Shannons's logarithmic
information measure $S$, regarded as a measure of {\it ignorance}
\cite{katz}, with  \be S=-k_B Tr \hat\rho \ln \hat\rho,
\label{entropy} \ee where $k_B$ is the Boltzmann constant, to be
set equal to unity from now on. Jaynes' basic  axiom or postulate
  reads: {\it the density operator that describes our system is
that provided by the MaxEnt principle}. One might argue that this
postulate explicitly assumes  Shannon's entropy logarithmic form,
but such an  statement can be refuted by pointing out that {\it
other} forms have been used in the literature to this effect with
great success \cite{tsallis,pla}. A second axiom that one needs,
however, is that $\hat \rho$
 depends explicitly only on expectation values and implicitly on some
hamiltonian.

 \nd Most interestingly, {\sf no reference to either i) equal a priori
probabilities
  or ii) equilibrium, needs to be made}. On IT grounds, equilibrium
  refers to the state of knowledge of the observer, it is not an
  intrinsic property of the system.
  Information theory is essentially concerned with epistemology:
  equilibrium means that one's
  knowledge is restricted to constants of the motion, so that one
  can forget about dynamics
  \cite{jaynes2}. The equilibrium notion plays no part whatsoever in
our
  considerations.

  \nd On Ockham's razor grounds, one might  argue that
Jaynes'  number of postulates is smaller that Gibbs'. In
particular, since no mention of ``equilibrium" is made,  the
associated theory has, at least potentially, a wider outreach than
that of Gibbs' \cite{jaynes2}. It is perhaps necessary to point
out, at this point, that an entirely different information theory
approach to non-equilibrium thermodynamics, based upon Fisher's
measure (a kind of ``Fisher-MaxEnt"), has recently been advanced
that exhibits definite advantages over both Gibbs' and (the
original) Jaynes' treatments \cite{nuestro,nuestro2,nuestro1}.

\nd Returning to Jaynes' approach, assume that we  deal with  a
system with Hamiltonian $\widehat H$. The system is characterized by
the set of operators $\{\widehat O_i\}\,\,\,(i=1,\ldots,M, \,\,{\rm
with}\,\, \widehat O_1\equiv\widehat H)$ in the sense that one is
supposed to know the expectation values of these operators. In other
words

\ba \label{constr}  \langle \widehat
O_i\rangle&=&a_i,\,\,(i=1,\ldots,M) \cr Tr[\hat \rho] &=&1,\ea
constitutes  our a priori information concerning the system. We wish
to find the appropriate, most unbiased $\hat \rho$ that reproduces
this amount of information and otherwise maximizes our ignorance.
The truth, all the truth, nothing but the truth \cite{katz}. Using
any other $\hat \rho$ is tantamount to {\it inventing} information
that we actually do not possess. Extremizing then (\ref{entropy})
subject to the constraints (\ref{constr}) leads to a density
operator of the form \cite{katz}

\be \hat\rho=\frac{e^{-\sum_i^M \lambda_i \widehat O_i}}{Z},
\label{rho1} \ee where the $\{\lambda_i\}$ is a set of Lagrange
multipliers ($\lambda_1=\beta$ is the inverse temperature) that
arise during the Lagrange process. The $\{\lambda_i\}$
 are associated with the above expectation values  that represent our
 foreknowledge \cite{katz}. The normalization factor in Eq.
(\ref{rho1}) \be Z=Tr\,[e^{-\sum_i^M \lambda_i \widehat O_i}] \ee is
the partition function \cite{katz}. The formalism allows for
arbitrary variations in the expectation values to be carried out.
Let us insist:  thermodynamics  {\it has been derived more than 40
years ago} from the IT formalism. If one relies on IT, it is clear
that, epistemologically, no additional thermodynamic notions are to
be presupposed in advance. Notice that predictions derived from the
IT-formalism amply exceed the scope of themes that conventional
thermodynamics is able to deal with \cite{jaynes,jaynes2}.

\nd In order to obtain the first law, the usual text-book approach
\cite{katz} analyzes   the variation of the internal energy $U$,
which is regarded as a functional of both i) the density operator
and ii) the Hamiltonian.

\section{Our present goal}

\nd We will show in this work that it is possible to perform a
{\it different} treatment that considers the internal energy as a
functional of {\it solely}  the density operator. According to
Ockham's razor, this way of handling the first law is to be
preferred to the traditional one, since now one  can dispense
with, in order to describe  the thermodynamic work $W$,  the two
following theoretical assumptions (or needs) of the traditional
approach \cite{katz}: \vskip 1mm

\subsection{Assumptions that will be no longer needed}
\begin{itemize}
\item reference to ``equilibrium"
\item   explicit dependence of the Hamiltonian
on some ``external" parameters $\chi$,
\item  recourse to the adiabatic theorem (AT) \cite{katz}.
\end{itemize}

\nd We can summarize the AT's contents as follows  \cite{katz}:
let us regard the Hamiltonian as depending on a parameter $\chi$
that evolves in time from an initial value $\chi_1$ to a final
value $\chi_2$, during a time-interval $\tau$, in the fashion \be
\label{AT}
\chi(t)=\chi_1+\frac{t}{\tau}\,(\chi_2-\chi_1);\,\,(\chi(0)=\chi_1,\,\,\chi(\tau)=\chi_2).
\ee \noindent In the limit of an exceedingly slow, physically
unrealizable $\chi$-change (i.e., for $\tau \rightarrow \infty$),
the time evolution that $\hat H(\chi(t))$ generates during the
temporal interval $[0,\tau]$ is such that, if the system is
represented at $t=0$ by an eigenstate of $\hat H(\chi_1)$, it will
be found in an eigenstate of $\hat H(\chi_2)$ at $t=\tau$. Indeed,
at any time $t$ in $[0,\tau]$ it will be encountered in an
eigenstate of $\hat H(\chi(t))$.
\section{A relation for $dU$}
\nd The possibility of eliminating recourse  to the AT should be a
 relief to many who consider it a somewhat
 suspect subterfuge: to our knowledge there has been no {\it direct,
assumption free}
 experimental confirmation of the claim that the final state
 of a system is really independent of its path from the initial
 state when work is performed \cite{pipard}. It goes without saying
that
the consequences of the first law are so well established that its
validity is beyond any reasonable  doubt \cite{pipard}.
Nevertheless, in the light of this situation a new proof of the
first law, specially one that avoids the awkward AT, should be of
great interest. \vskip 2mm \nd We now proceed to  derive a
relationship for $dU$. For this we need to deal with the internal
energy $U$, that is, with a special case ($i=1$) of (\ref{constr})

\be U=\langle \widehat O_1 \rangle\equiv\langle \widehat H
\rangle= Tr [\hat\rho\,\widehat H], \ee and consider variations
$\delta\hat\rho$ of the density operator, whose normalization
entails

\ba \label{norm} Tr[\delta\hat\rho]&=&\delta Tr[\hat\rho]= \delta
\hat 1=0 \cr  Tr [\delta\hat\rho\, \ln \hat\rho]&=&\delta Tr[
\hat\rho \,\ln \hat\rho]. \ea
We also have, of course, \be
\label{norm1}  Tr [\delta\hat\rho\,\widehat O_i]=\delta
\left<\widehat O_i\right>.\ee

\nd Thus, we confront finding (Cf. (\ref{norm1}))
 \be
dU \equiv  \delta \left<\widehat O_{i=1}\right>
=Tr[\delta\hat\rho\,\widehat H]. \label{dU} \ee

\nd Appropriate manipulation of  Eq. (\ref{rho1}) allows one now
to write  $\widehat H$ in the fashion

\be \ln\hat\rho+\ln Z= -\left[\beta\widehat H + \sum_{i>1}\lambda_i
\widehat O_i\right], \ee and thus

\ben \widehat H&=&-\frac{1}{\beta} \left(\ln\hat\rho+\ln
Z+\sum_{i>1}\lambda_i \widehat O_i\right);\cr \widehat
O_k&=&-\frac{1}{\lambda_k} \left(\ln\hat\rho+\ln Z+\sum_{i \ne
k}\lambda_i \widehat O_i\right), \een so that, replacing $\widehat
H$ into Eq. (\ref{dU}), and minding also (\ref{norm}-\ref{norm1}),
yields

\be dU=-\frac{1}{\beta}\,\delta \{Tr [\hat\rho \,\ln
\hat\rho]\}-\sum_{i>1}\,\frac{\lambda_i}{\beta}\,\delta
\left<\widehat O_i\right>. \label{dU2} \ee The first term in Eq.
(\ref{dU2}) is now to  be recast in terms of the entropy of the
system, as given by Eq. (\ref{entropy}), which leads one to

\be dU=TdS-\sum_{i>1}T\lambda_i\delta \left<\widehat O_i\right>, \ee
where $T=1/\beta$ is the temperature of the system. More generally,
one also has \be \delta \langle \widehat O_k \rangle
=\frac{dS}{\lambda_k}-\sum_{i \ne
k}\frac{\lambda_i}{\lambda_k}\delta \left<\widehat O_i\right>.
\label{indepe} \ee We make now the identification \be \label{Q} {\rm
for\,\,\, heat\,\,\,(Q)\,\,\, change:}\,\,\, d'Q=TdS, \ee and  we
arrive at the promised relationship for $dU$

\be \label{mistery} dU=d'Q+  dX, \ee with
$X=-T\sum_{i>1}\,\lambda_i\,\delta \left<\widehat O_i\right>.$

\section{What is $X$?}

\nd The derivation of (\ref{mistery}) is straightforward. We are
left with the interpretation of $X$. Let us delve a little longer on
 the meaning of (\ref{mistery}). We have assumed that our a priori
 information has slightly changed:
 \ba \label{constr1} {\rm From}\,\,\, \langle \widehat
O_i\rangle&=&a_i\,\,\,{\rm to}\cr
 \langle \widehat O_i\rangle +
\delta\langle \widehat O_i\rangle&=&a_i+\delta
a_i;\,\,\,\,\,\,(i=1,\ldots,M). \ea Necessarily then, the MaxEnt
methodology yields a new density operator $\hat \rho + \delta \hat
\rho$, which, of course, entails in turn a change in the Lagrange
multipliers \be \label{constr2} {\rm From}\,\,\,
\lambda_i\,\,\,{\rm to} \,\,\,\, \lambda_i +
\delta\lambda_i;\,\,\,\,\,(i=1,\ldots,M). \ee  The essential {\sf
IT-content} of the first law (Cf. Eq. (\ref{mistery})) is that i)
the $\delta\langle \widehat O_i\rangle$ are not independent
quantities (Cf. Eq. (\ref{indepe})) and ii) they can be expressed
solely in terms of  the $\lambda_i$.  Of course, if one wishes to
predict the value of $\langle \widehat A\rangle$, an operator not
included in the set $\{ \widehat O_i \}$, one would need the
$\delta\lambda_i$.  \vskip 3mm

\nd If we call  the differential of work (dW), effected at
temperature $T$,

\be \label{W} dW=-\sum_{i>1}T\lambda_i\delta \left<\widehat
O_i\right>, \ee we obviously obtain, without further ado, the first
law of thermodynamics in the fashion (\ref{mistery}),  where heat
($d'Q$) and work ($dW$) terms acquire their traditional aspect. If
we did not accept, for whatever the reason, the interpretation
(\ref{W}), we could not avoid  the conclusion that the difference
$dU-d'Q$ has {\bf two} forms: one of them follows from
(\ref{mistery}) and  {\it always} holds. In some particular
instances, we would, in addition, have the conventional first law.
On Ockham grounds, the first alternative, namely, Eq.
(\ref{mistery}), is clearly preferable. Consider the simple
classical example posed by a probability distribution $f(\tau)$ in
phase-space (volume element $d\tau$), with two constraints: \ben
\label{examp} f(\tau) &=& Z^{-1} \exp{\big(-[\beta H(\tau)+p \,\beta
\phi(\tau)]\big)}\cr \int\,d\tau\,f(\tau)\,H(\tau)&=&U; \,\,\,
\int\,d\tau\,f(\tau)\,\phi(\tau)= V\cr (\rm
V&\equiv&\,Volume;\,\,\,\,\,p \equiv\,pressure), \een where
application of Eq. (\ref{mistery}) immediately yields \be dU\,=\,
T\,dS-p\,\,dV.\ee
\section{Discussion and conclusions}

\nd We have shown that, within Jaynes' information theory context,
one may derive thermodynamics' first law without appeal  to the
adiabatic theorem \cite{katz} or to a explicit dependence of the
pertinent Hamiltonian on hypothetical external parameters. This
agrees with both Ockham's razor and the Jaynes' philosophy
\cite{jaynes2}. Thus we avoid the slightly paradoxical
contradiction between simultaneously stating
\begin{itemize} \item on the one hand,  that $\hat \rho$ contains {\it
all} the available information concerning the system, and, \item on
the other one, needing to add, to the theoretical description,
putative infinitely slowly varying external parameters to obtain the
first law.
\end{itemize} There is no need to
invoke the adiabatic theorem because, interestingly enough, the
formalism itself demands that the process be undertaken at a
constant temperature $T$ (Cf. Eqs. (\ref{Q})-(\ref{W})) that
arises automatically in the constrained Lagrange extremization.

\nd For  a system characterized by the set of operators
$[\{\widehat O_i\}\,\,\,(i=1,\ldots,M);\,\, \widehat O_1 \equiv
\widehat H]$, in the sense that we know a priori the pertinent
expectation values \cite{katz}, we have here shown that the Jaynes
treatment, in the present light, implies that  {\it work is
represented by {\it changes} in the expectation values}
$[\{\langle \widehat O_i \rangle \};\,\,\,(i=2,\ldots,M)]$. These
constituted part of our prior knowledge. If a posteriori we
encounter changes, this entails that work has been performed, on
or by the system.


\begin{thebibliography}{99}
\bibitem{nuestro1} The Jaynes-with-Shannon MaxEnt approach (J/S) yields
only exponential distributions, which lacks a certain degree of
``flexibility", whereas the  Jaynes-with-Fisher MaxEnt treatment
(J/F) \cite{nuestro} leads to partial differential equations,
whose solutions are of quite different forms. The concomitant J/F
solutions, in turn,  allow one to easily ascertain whether one has
found an equilibrium or a non-equlibrium distribution
\cite{nuestro}.
\end{thebibliography}
\end{document}